\documentstyle[preprint,aps,epsfig]{revtex}

\newcommand{\beq}{\begin{equation}}
\newcommand{\eeq}{\end{equation}}
\newcommand{\beqs}{\begin{eqnarray}}
\newcommand{\eeqs}{\end{eqnarray}}
\newcommand{\bary}{\begin{array}}
\newcommand{\eary}{\end{array}}

\newcommand{\figpos}{t}        % figure position may be h,t,b or p
\newcommand{\andthis}{~~~~~~\mbox{and}~~~~~~}

\def\gsim{\ \rlap{\raise 3pt \hbox{$>$}}{\lower 3pt \hbox{$\sim$}}\ }
\def\lsim{\ \rlap{\raise 3pt \hbox{$<$}}{\lower 3pt \hbox{$\sim$}}\ }

\def\putFig#1#2#3#4#5 % #1=filename, 
{
 \newpage 
 \begin{figure}[\figpos]
 \LARGE
 $y(x)$ 
 \begin{center}
 \mbox{\epsfig{figure=#1.eps,angle=0,width=#3cm,height=#4cm}}
 \end{center}
 \hspace{#5cm}
 $x$
 \vspace{1cm}
 \caption{#2}
 \label{#1}
 \end{figure}
}

\begin{document}
%%%%%%%%%%%%%%%%

\title{\LARGE Generic Smooth Connection Functions \\
       A New Analytic Approach to Interpolation}

\author{Alex Alon$^a$ and Sven Bergmann$^b$ \\
	\small \it $^a$CEO, Blurbusters Co., Masarik, 
        Tel Aviv 64351, Israel \\
        \small \it $^b$Department of Particle Physics\footnote{
         From 16.5.2001: Department of Molecular Genetics},
        Weizmann Institute of Science,
        Rehovot 76100, Israel \\
        \small \rm e-mail: {\tt Sven.Bergmann@weizmann.ac.il}} 

\maketitle
\vspace{1.5cm}

\begin{abstract}%
  We present a generic solution to the fundamental problem of how to
  connect two points in a plane by a smooth curve that goes through
  these points with a given slope. The smoothness of any curve depends
  both on its curvature and its length. The smoothest curves
  correspond to a particular compromise between minimal curvature and
  minimal length. They can be described by a class of functions that
  satisfy certain boundary conditions and minimize a weight
  functional. The value of this functional is given essentially by the
  average of the curvature raised to some power $\nu$ times the length
  of the curve. The parameter $\nu$ determines the importance of
  minimal curvature with respect to minimal length.  In order to find
  the functions that obtain the minimal weight, we use extensively
  notions that are well-known in classical mechanics.  The
  minimization of the weight functional via the Euler-Lagrange
  formalism leads to a highly non-trivial differential equation.
  Using the symmetries of the problem it is possible to find conserved
  quantities, that help to simplify the problem to a level where the
  solution functions can be written in a closed form for any given
  $\nu$. Applying the appropriate coordinate transformation to these
  solutions allows to adjust them to all possible boundary conditions.
\end{abstract}%

\newpage
 
%=====================
\section{Introduction}
%=====================

%-----------------------
\subsection{The problem}
\label{problem}
%-----------------------

Consider two points in a plane, each associated with a ray pointing in
some direction. The problem we address in this paper is how to connect
these points in a ``smooth'' way. That is we are looking for a curve
that satisfies the following three conditions:
\begin{enumerate}
\item The curve goes through both points.
\item The associated ray at each point is a tangent to the curve.
\item Between the two points the curve follows some {\it optimal}
  path, which is a compromise between minimal curvature and minimal
  length.
\end{enumerate}
The first two requirements constitute the boundary conditions of the
problem. The third condition defines what we mean by ``smooth''.

%----------------------------------
\subsection{Motivation and Outline}
%----------------------------------

\noindent
A solution to the problem outlined above has many obvious
applications. For example it could be used to determine the ideal
shape of a road to be built between two points where its direction is
predetermined (e.g. by two bridges). In fact {\it any\/} interpolation
problem that has been reduced to the task of connecting a sample of
data points where the slope is fixed can be solved using the
elementary solution of connecting smoothly two points.  If the
original sample only consists of points, the respective slopes can be
determined, for example, by taking the slope of the line through the
neighboring points, or by some different, more sophisticated method.
Evidently the quality of interpolation curves is crucial to all fields
that deal with numerical data from applied sciences to economics.  It
is interesting to note that while there exist many interpolation
schemes, most of them rely on simple functions, like polynomials, that
in general do not give the best interpolation. The aim of this work is
to establish first a criterion for the quality of the interpolation
and then to investigate quantitatively which are the optimal
interpolating curves.

As we shall see in the following, the problem outlined above
translates into a well-defined mathematical exercise, which is
interesting by itself.  In fact we feel that our solution to the
problem, relying heavily on notions well known in physics, has also a
pedagogical value.  It serves as an example of some fundamental
physics principles in the context of a very intuitive and visual, yet
non-trivial problem.

Since the problem is defined as a geometrical task, its mathematical
formulation obviously does not depend on time.  Instead we take one
space-direction ($x$) as the variable of integration, and the other
space-direction ($y$) to describe the solution function.  While this
description is not manifestly invariant under rotations and
translations in the $x-y$ plane, it allows us to define an {\it
  action\/} functional that assigns a weight to any function (that
complies with the boundary conditions) according to its
``smoothness''. It is given essentially by the average of the inverse
curvature radius, to some power $\nu$, times the length of the curve.
This action is minimized by a class of functions, each describing the
smoothest curve for a particular choice of the parameter $\nu$ that
specifies the relative importance of minimal curvature with respect to
minimal length. These functions are the solutions of the {\it
  Euler-Lagrange equation}, which turns out to be a complicated,
non-linear third order differential equation.  Solving this equation
can be facilitated immensely by taking advantage of the symmetries of
the problem. We compute the {\it linear\/} and {\it angular momenta\/}
that follow from {\it Noether's theorem\/} due to the {\it
  translational\/} and {\it rotational invariance}.  The equation of
motion is also {\it scale invariant}, but there exists no conserved
charge corresponding to scaling, since the Lagrangian and other
dimensionful variables do change under scaling transformations.  The
problem is solved explicitly for a particular choice of the momenta.
Applying the appropriate coordinate transformation to this solution
allows to adjust it to all possible boundary conditions.

%==================
\section{Formalism}
\label{formalism}
%==================

In order to translate the problem outlined in Section~\ref{problem}
into a mathematical one we have to introduce some notations: Let us
give ``names'' to the two points: We shall refer to them as the
initial point $P_i$ and the final point $P_f$. Which one is which is
arbitrary, but when introducing coordinates
\beq \label{coordinates}
P_i: (x_i,y_i) \andthis P_f: (x_f,y_f) \,,
\eeq
we demand that $x_i < x_f$. The coordinates $(x,y)$ are a set of two
real numbers specifying any point in the plane. We use simple
Euclidean geometry. The associated rays at $P_i$ and $P_f$ are
described by their inclination angles, $\alpha_i$ and $\alpha_f$, with
respect to the vector pointing from $P_i$ to $P_f$
(c.f.~Fig.~\ref{SCF}).

We call the function $y(x)$ that satisfies the three conditions in
Section~\ref{problem} the ``smooth connection function'' (SCF).  Our
task is to determine this function.  The boundary conditions defined
in Section~\ref{problem} are:
\beqs
y(x_i)  &=& y_i ~~~~~~~~~~~~~~~\andthis~ 
y(x_f)   =  y_f \,, \label{boundary1} \\
y'(x_i) &=& \tan(\alpha_i+\alpha_0) ~\andthis \,
y'(x_f)  =  \tan(\alpha_f+\alpha_0) \,, \label{boundary2}
\eeqs
where the prime denotes a derivative with respect to $x$ and
$\alpha_0$ is the angle between the positive $x$-axis and the vector
pointing from $P_i$ to $P_f$.

In order to include the third and crucial condition in
Section~\ref{problem} we need to incorporate both the curvature and
the length of the curve. The {\it local curvature\/} at a given point
on the curve is defined as the inverse radius $(1/r)$ of the circle
that coincides with the curve in the infinitesimal vicinity of the
point. It is given in terms of the first and second derivative of the
function $y(x)$ by:
\beq \label{curvature}
{1 \over r(x)} = {|y''(x)| \over [1+y'(x)^2]^{3/2}} \,.
\eeq
To show this relation it is enough to verify that the curves that
describes a circle, i.e. $y(x) = \pm \sqrt{r^2 - (x-x_0)^2} + y_0$,
where $r$ is the radius and $(x_0,y_0)$ denotes the center of the
circle, give $r(x) = r$. The length of an infinitesimal piece of the
curve at $x$ is given by
\beq \label{lengthelements}
ds(x) = \sqrt{1+y'(x)^2} \, dx \,.
\eeq
We define now the following functional:
\beq \label{action}
\tilde {\cal S}[y(x)] \equiv 
\int_{x_i}^{x_f} \left({1 \over r(x)}\right)^\nu \, ds(x) = 
\int_{x_i}^{x_f} {|y''(x)|^\nu \over 
                  \left[1+y'(x)^2\right]^{3\nu-1 \over 2}} \,dx \,.
\eeq
We shall refer to $\tilde {\cal S}[y(x)]$ as the {\it action}.  For a
given function $y(x)$ it returns the weighted sum of $ds(x)$ where $x$
goes from $x_i$ to $x_f$. The weight at position $x$ is given by the
local curvature $1/r(x)$, defined in eq.~(\ref{curvature}), raised to
some power $\nu$. A priori this parameter is an arbitrary real number
that determines how to choose the compromise between minimal curvature
and minimal length.  For example, for $\nu=0$ the curvature does not
play a role at all and $\tilde {\cal S}[y(x)]_{\nu=0}$ is just the
length of the curve.  For $|\nu| \to \infty$ the situation is exactly
the opposite, since $\tilde {\cal S}[y(x)]_{|\nu| \to \infty} \simeq
\int (1/r)^\nu dx$ only depends on the curvature in this case.  A
special situation arises for $\nu=1$, where $\tilde {\cal S}[y(x)] =
\int_{x_i}^{x_f} (ds/r) = \int_{x_i}^{x_f} d\alpha =
\alpha_f-\alpha_i$. Since the action gives just the difference between
the inclination angles at the boundary, independent of the particular
choice of $y(x)$, it is impossible to determine the optimal path for
$\nu=1$ and we exclude this case from our subsequent discussion.

In order to find the optimal curve we have to minimize the action
under the boundary conditions in eq.~(\ref{boundary1}) and
eq.~(\ref{boundary2}).  Before we continue with our analysis we would
like to give a concrete example illustrating the relevance of the
action functional. Consider a spacecraft that has been launched to
explore some distant planets and which is flying with constant
velocity $v$ through the interstellar space. When approaching a
certain planet, one has to adjust carefully the trajectory of the
spaceship, say to enter the atmosphere of the planet at a precise
angle or to use its gravitational field to accelerate the spacecraft
to its next destination. Suppose that for this purpose the spaceship
has thrusters that exert a force $F$ perpendicular to the direction of
motion. Then at any given time $t$ between the ignition (at $t_i$) and
the end of the manoeuvre (at $t_f$) the instantaneous curvature radius
$r(t)$ is inversely proportional to $F$. Let us assume that the fuel
consumption per unit time, ${\cal W}$, is governed by some potential
law, i.e.  ${\cal W} \propto F^\nu \propto (1/r)^\nu$, where $\nu$ is
an empirical parameter. Then the total fuel consumption, ${\cal C} =
\int_{t_i}^{t_f} {\cal W} dt$, is proportional to the action in
eq.~(\ref{action}), since $dt=ds/v$.  Obviously minimizing ${\cal C}$
is crucial. Once we have found the solution for the trajectory we can
obtain the curvature radius as a function of $t$ which indicates how
much fuel should be burned at each point.

The integrand in eq.~(\ref{action}), usually referred to as the {\it
  Lagrangian},
\beq
\tilde {\cal L}(y',y'') \equiv 
{|y''(x)|^\nu \over \left[1+y'(x)^2\right]^{3\nu-1 \over 2}} 
\eeq
only depends on $y'(x)$ and $y''(x)$, while there is no explicit
dependence on $y(x)$ and $x$. Thus $y$ and $x$ enter the action
$\tilde {\cal S}[y(x)]$ only via the boundary conditions.  From
eq.~(\ref{boundary1}) it follows that
\beq
y_f - y_i = y(x_f) - y(x_i) = \int_{x_i}^{x_f} y'(x) = constant \,.
\eeq
This constraint can be absorbed into the action functional by
introducing a Lagrange multiplier $\lambda$, resulting in a new action
\beq \label{action2}
{\cal S}[y'(x)] \equiv 
\tilde {\cal S}[y'(x)] + \lambda \int_{x_i}^{x_f} y'(x) =
\int_{x_i}^{x_f} 
\left[ \tilde {\cal L} + \lambda \, y'(x) \right] \, dx \,.
\eeq
Since there is no reference to $y(x)$ anymore, but only to its first
and second derivative, we can change variables,
\beq
q(x) = y'(x) \andthis q'(x) = y''(x) \,,
\eeq
and write the Lagrangian corresponding to the action ${\cal S}[q(x)]$ in  
eq.~(\ref{action2}) as
\beq \label{Lagrangian}
{\cal L}(q,q') \equiv \tilde {\cal L}(q,q') + \lambda \, q(x) =
{|q'|^\nu \over \left(1+q^2\right)^{3\nu-1 \over 2}} + 
 \lambda \, q \,.
\eeq 
Now, to minimize $\tilde {\cal S}[y(x)]$ under the boundary condition
in eq.~(\ref{boundary1}) is equivalent to the minimization of ${\cal
  S}[q(x)]$. A necessary condition for ${\cal S}[q(x)]$ to be extremal
is that its first functional derivative $\delta{\cal S}[q(x)]/\delta
q(x)$ vanishes.  Since ${\cal S}[q(x)]$ only depends on $q$ and $q'$,
this is a well-known problem. Its solution is given by the {\it
  Euler-Lagrange equation\/}:
\beq \label{EulerLagr}
{\partial {\cal L} \over \partial q} - 
{d \over dx} \, {\partial {\cal L} \over \partial q'} = 0 \,.
\eeq
Computing
\beqs
{\partial {\cal L} \over \partial q} &=& 
{(1-3\nu) |q'|^\nu q \over 
 \left(1+q^2\right)^{3\nu+1 \over 2}} + \lambda \,, \label{dLdq} \\
{\partial {\cal L} \over \partial q'} &=& 
{\nu \, \sigma |q'|^{\nu-1} \over 
 \left(1+q^2\right)^{3\nu-1 \over 2}} \,, \label{dLdqp} \\
{d \over dx} \, {\partial {\cal L} \over \partial q'} &=&
{\nu (\nu-1) |q'|^{\nu-2} q'' \over \left(1+q^2\right)^{3\nu-1 \over 2}} +
{\nu (1 - 3\nu) |q'|^\nu q \over \left(1+q^2\right)^{3\nu+1 \over 2}} \,,
\eeqs
where $\sigma \equiv {\rm sign}(q')$, we obtain 
\beq \label{EOMlambda}
\lambda = {(\nu-1) |q'|^{\nu-2} \over \left(1+q^2\right)^{3\nu+1 \over 2}}
 \left[\nu (1+q^2) q'' + (1-3\nu) (q')^2 q \right] \,.  
\eeq
The above equation still depends on the parameter $\lambda$. We can
eliminate this dependence by differentiating eq.~(\ref{EOMlambda})
with respect to $x$ yielding the following {\it equation of motion\/}
(EOM)
\beqs \label{EOM}
0 = {(\nu-1) |q'|^{\nu-3} \over \left(1+q^2\right)^{3\nu+3 \over 2}}
  &~& \left\{ (1-3\nu) \left[(1-3\nu q^2) (q')^4 +
              2\nu q (1+q^2) (q')^2 q'' \right] \right. + \nonumber \\
  &~& \left. ~~\nu (1+q^2)^2 \left[(\nu-2) (q'')^2 + q' q''' \right]
 \right\} \,.
\eeqs
For all functions $q(x)$ that solve the above EOM the action ${\cal
  S}[q(x)]$ in eq.~(\ref{action2}) is {\it stationary}, i.e.
$\delta{\cal S}[q(x)]/\delta q(x)=0$. However, in order to {\it
  minimize \/} ${\cal S}[q(x)]$ also the so-called Legendre condition,
\beq \label{Legendre}
0 < {\partial^2 {\cal L} \over (\partial q')^2} = 
{\nu (\nu-1) |q'|^{\nu-2} \over 
 \left(1+q^2\right)^{3\nu-1 \over 2}} \,,
\eeq
has to be satisfied. This condition arises from the second functional
variation of the action with respect to $q(x)$. Note that unlike for
the minimization of usual functions it is only a {\it necessary\/}
condition for a minimum of ${\cal S}[q(x)]$. This is essentially
because a small variation in $q$ does not necessarily imply a small
variation in $q'$. However in our case the sign of ${\partial^2 {\cal
    L} / (\partial q')^2}$ only depends on the prefactor $\nu
(\nu-1)$, but is independent of $q$ and $q'$ even if they are taken as
{\it independent\/} variables. Such a situation is referred to as a
``regular problem'' and it implies that the solutions of the EOM
indeed give rise to a local minimum of the action if $\nu<0$ or
$\nu>1$. Conversely for $0<\nu<1$ one obtains a local maximum. Thus we
rule out any $0 < \nu \le 1$ from our analysis and we shall only
consider $\nu \le 0$ and $\nu>1$ from now on.

Even though we have managed to translate the problem at hand into a
differential equation, solving this equation analytically seems like a
formidable task given that eq.~(\ref{EOM}) is third order in $q$ and
non-linear. However, the integrated form in eq.~(\ref{EOMlambda})
gives us a hint of how to proceed. The fact that $\lambda$ is a
constant is related to the symmetries inherent to the problem.
Studying these symmetries systematically one can simplify the problem
significantly and find its solutions.

%===================
\section{Symmetries}
%===================

Let us examine more carefully the problem with respect to its
intrinsic symmetries.  The important point to realize is that when
introducing the coordinates in eq.~(\ref{coordinates}) we have made an
explicit choice of
\begin{itemize}
\item where to define the origin, 
\item how to orientate the axes and 
\item what units of length to use. 
\end{itemize}
These choices are arbitrary and once we have found a solution to our
problem we can redefine our coordinate system. Assume that $y(x)$
solves the EOM in eq.~(\ref{EOM}) in some coordinate system $\Sigma$
and that $\tilde y(\tilde x)$ is obtained from $y(x)$ by a
transformation to a different coordinate system $\tilde \Sigma$\,,
\beq \label{transf}
\pmatrix{x \cr y} \to  
\pmatrix{\tilde x \cr \tilde y} = 
 {\bf R} \pmatrix{x \cr y} + \pmatrix{x_0 \cr y_0} \,,
\eeq
where
\beq \label{Rmat}
{\bf R} \equiv \pmatrix{\cos \theta & -\sin \theta \cr
                        \sin \theta &  \cos \theta}
\eeq
describes a {\it rotation\/} by an angle $\theta$, and $x_0$ and $y_0$
parameterize a {\it translation} in the $x$-direction and in the
$y$-direction, respectively. The action in eq.~(\ref{action}) is
defined in terms the curvature radius $r$ [c.f. eq.~(\ref{curvature})]
and the length element $ds$ in eq.~(\ref{lengthelements}). Both
quantities are manifestly invariant under rotations and translations.
Therefore it is clear that the coordinate transformations in
eq.~(\ref{transf}) do not change the action nor the EOM derived from
it. However the boundary conditions obviously have to change under
$\Sigma \to \tilde \Sigma$.

Let us consider the infinitesimal change of the coordinate $y$ {\it at
  a given $x$},
\beq
\delta y(x) \equiv \tilde y(x) - y(x) \,,
\eeq
for the three types of coordinate transformations. For a translation
in the $y$-direction
\beq
y_0    = \epsilon_y \ll 1 \,,~~~~~
x_0    = 0 \,,~~~~~
\theta = 0 \,,
\eeq
the answer is trivial. The change in $y$ is simply
\beq \label{dely-transly}
\delta y(x,\epsilon_y) = \epsilon_y \,,
\eeq
implying that the changes in all the derivatives of $y(x)$ vanish:
\beqs \label{delq-transly}
\delta q(x,\epsilon_y)   &=& 0  \,, \nonumber \\
\delta q'(x,\epsilon_y)  &=& 0  \,, \\
\delta q''(x,\epsilon_y) &=& 0  \,. \nonumber
\eeqs
However, a translation in the $x$-direction
\beq
y_0    = 0 \,,~~~~~
x_0    = \epsilon_x \ll 1 \,,~~~~~
\theta = 0 \,,
\eeq
is a bit more tricky. First, note that
\beq
\tilde y(\tilde x) = y(x) \,.
\eeq
This expresses merely the fact that the coordinate transformation in
eq.~(\ref{transf}) leaves the functional behavior of $y(x)$ invariant.
It does not make a difference whether we refer to the function by
$y(x)$ in the coordinate system $\Sigma$ or $\tilde y(\tilde x)$ in
the system $\tilde \Sigma$. Then a Taylor expansion of $\tilde
y(\tilde x)$ around $x$ gives
\beq 
\tilde y(\tilde x) = % y(x) = 
\tilde y(x) + \left.{\partial y \over \partial x}\right|_x 
 \, \epsilon_x \,,
\eeq
implying that
\beq \label{dely-translx}
\delta y(x,\epsilon_x) = 
-\left.{\partial y \over \partial x}\right|_x \, \epsilon_x \,.
\eeq
From eq.~(\ref{dely-translx}) it follows that the corresponding
changes in the derivatives of $y(x)$ are given by
\beqs \label{delq-translx}
\delta q(x,\epsilon_x)   &=& -q'(x)   \, \epsilon_x   \,, \nonumber \\
\delta q'(x,\epsilon_x)  &=& -q''(x)  \, \epsilon_x  \,, \\
\delta q''(x,\epsilon_x) &=& -q'''(x) \, \epsilon_x \,. \nonumber
\eeqs
Finally, an infinitesimal rotation
\beq
y_0    = 0 \,,~~~~~
x_0    = 0 \,,~~~~~
\theta = \epsilon_\theta \,,
\eeq
involves both a change in $x$ by $-y \epsilon_\theta$ and in $y$ by $x
\epsilon_\theta$. Together this results into a change of $y(x)$ by:
\beq \label{changey-rot}
\delta y(x,\epsilon_\theta) =
\left(x + \left.{\partial y \over \partial x}\right|_x  y \right) 
 \, \epsilon_\theta \,.
\eeq
From eq.~(\ref{changey-rot}) it follows that the changes in $q$, $q'$
and $q''$ for an infinitesimal rotation are given by
\beqs \label{delq-rot}
\delta q(x,\epsilon_\theta)   &=& 
 (1 + q^2 + y q') \, \epsilon_\theta \,, \nonumber \\
\delta q'(x,\epsilon_\theta)  &=& 
 (3 q q' + y q'') \, \epsilon_\theta \,, \\
\delta q''(x,\epsilon_\theta) &=& 
 (3 (q')^2 + 4 q q'' + y q''') \, \epsilon_\theta \nonumber \,.
\eeqs

Let us now investigate the effect of the coordinate transformations in
eq.~(\ref{transf}) on the Lagrangian ${\cal L}$ in
eq.~(\ref{Lagrangian}).  In general an infinitesimal transformation
parameterized by $\epsilon \ll 1$ will induce a change
\beq
\delta {\cal L} \equiv {\cal L}(\tilde q, \tilde q') - {\cal L}(q, q') \,. 
\eeq
Provided that ${\cal L}$ has no explicit dependence on $x$, a Taylor
expansion of ${\cal L}(\tilde q, \tilde q')$ around $q$ gives
\beqs
\delta {\cal L} &=& {\partial {\cal L} \over \partial q}  \delta q + 
             {\partial {\cal L} \over \partial q'} \delta q' \\
&=& 
{d \over dx} \left({\partial {\cal L} \over \partial q'} \delta q \right) +
\left({\partial {\cal L} \over \partial q} - 
{d \over dx} \, {\partial {\cal L} \over \partial q'} \right) \delta q \,.
\label{expandL} 
\eeqs
The second term in eq.~(\ref{expandL}) vanishes for $q(x)$ that
satisfy the equation of motion~(\ref{EulerLagr}).

Any transformation that changes the action ${\cal S}[q(x)]$ in
eq.~(\ref{action2}) at most by constant is called a {\it symmetry
  transformation}. Such a transformation does not change the extremal
condition for the action and therefore leaves the EOM invariant. (Note
however that not all transformations that leave the EOM unchanged are
symmetry transformations according to our definition.)  Then, the
corresponding Lagrangian could change only by a total derivative:
\beq
\delta {\cal L} = \epsilon {dF \over dx} \,,
\eeq
where $F(q,q')$ is a function of $q$ and $q'$.  Equating the two
results for $\delta {\cal L}$ we find that
\beq \label{dF}
\epsilon {dF \over dx} = 
{d \over dx} \left({\partial {\cal L} \over \partial q'} 
 \delta q \right) \,.
\eeq
We stress that the expression on the left-hand side is correct for a
symmetry transformation of {\it any} function $y(x)$, while the
right-hand side is only valid for a transformation of $y(x)$ that
solves the EOM. From eq.~(\ref{dF}) it follows that the {\it conserved
  charge} defined as
\beq
Q = F - {\partial {\cal L} \over \partial q'} {\delta q \over \epsilon}
\eeq
is a constant with respect to $x$, i.e. $dQ/dx=0$. Consequently it has
one specific value for all points on a particular solution of the
EOM.  Obviously $Q$ can only be defined up to a constant. The above
argument, that every continuous symmetry transformation implies a
conserved charge is known as the {\it Noether theorem}.

To be explicit let us apply it to a translation in the $x$ direction.
Using eq.~(\ref{delq-translx}) it follows that
\beqs
\delta {\cal L}(x,\epsilon_x) &=& 
- \left({\partial {\cal L} \over \partial q}  q' + 
        {\partial {\cal L} \over \partial q'} q'' \right) 
\, \epsilon_x \\
&=&
- {d{\cal L} \over dx} \epsilon_x \,.
\eeqs
This means that (up to a constant) we can identify $F_x = -{\cal L}$.
Then it follows that the conserved charge corresponding to
translations in the $x$ direction is given by:
\beqs \label{Px}
P_x &=&  {\partial {\cal L} \over \partial q'} q' - {\cal L} 
\nonumber \\
&=& {{(\nu-1) |q'|^\nu} \over {\left(1+q^2 \right)^{3\nu-1
      \over 2}}} - \lambda \, q \,.
\eeqs
We call $P_x$ the {\it conserved momentum in the $x$ direction}.  In
fact it is nothing more than the Legendre transformation of the
Lagrangian with respect to $q'$. (If the variable of integration in
eq.~(\ref{action2}) had been the time $t$ rather the coordinate $x$
then the equivalent Legendre transformation of the Lagrangian with
respect to $dq/dt$ is called the {\it Hamiltonian} and the charge
related to time invariance is the {\it energy}.)

One might guess that a similar argument for translations in the $y$
direction should give another conserved charge $P_y$, which is the
{\it conserved momentum in the $y$ direction}. However a change in the
variable $y(x)$ as in eq.~(\ref{dely-transly}) has no effect on $q(x)$
and $q'(x)$, see eq.~(\ref{delq-transly}), implying trivially that
\beq
\delta {\cal L}(x,\epsilon_y) = 0 \,.
\eeq
Thus Noether's theorem does not help in this case to derive the
conserved charge. However we have already encountered another
conserved quantity, which could serve as a candidate for $P_y$.  In
order to absorb the boundary condition into the action we introduced
the Lagrange multiplier $\lambda$, which intuitively is related to
changes in $y$, c.f. eq.~(\ref{action2}). Eq.~(\ref{EOMlambda}) states
that $\lambda$ equals to a complicated function of $q$, $q'$ and $q''$
for all $x$. Therefore this function is a constant of motion.  To
prove that indeed $P_y=\lambda$ is nontrivial and we will show this
after discussing the conserved charge related to the rotations in
eq.~(\ref{transf}).

Using eqs.~(\ref{dLdq}) and~(\ref{dLdqp}) and the changes in $q$ and
$q'$ under rotations according to eq.~(\ref{delq-rot}) the change in
the Lagrangian under an infinitesimal rotation is
\beq
\delta {\cal L}(x,\epsilon_\theta) = 
\left\{
\left[{(1-3\nu) |q'|^\nu q \over \left(1+q^2\right)^{3\nu+1 \over 2}}
  + \lambda \right] \, (1 + q^2 + y q') +
\left[{\nu |q'|^{\nu-1} \over \left(1+q^2\right)^{3\nu-1 \over 2}}
  \right] \, (3 q q' + y q'')
\right\} \epsilon_\theta \,.
\eeq
It is not difficult to check that this can be rewritten as a total
derivative,
\beq
\delta {\cal L}(x,\epsilon_\theta) = {d \over dx} 
 \left[{y |q'|^\nu \over \left(1+q^2\right)^{3\nu-1 \over 2}} +
 \lambda (x + y q) \right] \epsilon_\theta \,.
\eeq
Thus for infinitesimal rotations the function $F$ can be identified to
be (up to a constant)
\beq
F_\theta  = y {\cal L} + x \lambda \,.
\eeq
Then the charge corresponding to the rotation symmetry, which is
called the {\it total angular momentum}, is given by
\beqs
J &=& F_\theta - 
 {\partial {\cal L} \over \partial q'} 
 {\delta q(x,\epsilon_\theta) \over \epsilon_\theta} \\
&=& y {\cal L} + x \lambda - 
 {\nu \, \sigma |q'|^{\nu-1} \over 
  \left(1+q^2\right)^{3\nu-1 \over 2}}
 \, (1 + q^2 + y q') \\
&=& \left(x P_y - y P_x \right) + S \label{J} \,.
\eeqs
In the last step we have separated the total angular momentum into the
orbital contribution
\beq
L = x P_y - y P_x
\eeq
and the remaining term
\beq
S = -{\nu \, \sigma |q'|^{\nu-1} \over 
      \left(1+q^2\right)^{3\nu-3 \over 2}}
  = -{\nu \, \sigma \over [r(x)]^{\nu-1}} \,,
\eeq
which we call the {\it spin} or {\it intrinsic angular momentum}.  The
{\it orbital angular momentum} would be the sole contribution if
$y(x)$ is a straight line (as is the case for $\nu=0$ or if $q'=0$). A
non-vanishing spin $S$ arises for all the other curves due to their
curvature. Note that it is sensitive to the sign $\sigma$ of $q'$.

We would like to come back now to our claim that $P_y$ coincides with
$\lambda$ defined in eq.~(\ref{EOMlambda}).  To this end let us
compute the change in $P_i~~(i=x,y)$ induced by rotations. In general
\beq \label{Pi}
\delta P_i(x,\epsilon_\theta)  = 
\left[{\partial P_i \over \partial q} (1+q^2) +
      {\partial P_i \over \partial q'} (3q q') +
      {\partial P_i \over \partial q''} (3[q]^2 + 4 q q'') \right]
\epsilon_\theta \,,
\eeq
where we used the results in eq.~(\ref{delq-rot}) and the fact that
the terms proportional to $y$ add up to the total derivative of $P_i$:
\beq
y \left[{\partial P_i \over \partial q}   q' +
        {\partial P_i \over \partial q'}  q'' +
        {\partial P_i \over \partial q''} q''' \right]
\epsilon_\theta = 
y {dP_i \over dx} \epsilon_\theta = 0 \,.
\eeq
Assuming that indeed $P_y = \lambda$, a somewhat tedious but 
straight-forward calculation of $\delta P_y$ according to
eq.~(\ref{Pi}) yields
\beq
\delta P_y(x,\epsilon_\theta) = 
{(\nu-1) |q'|^{\nu-2} \over \left(1+q^2\right)^{3\nu+1 \over 2}}
\left[(1+3\nu q^2) (q')^2 - \nu q (1+q^2) q'' \right] \epsilon_\theta
= P_x \epsilon_\theta \,. 
\eeq
Similarly using eq.~(\ref{Pi}) one shows that $\delta
P_x(x,\epsilon_\theta) = - P_y \epsilon_\theta$.  From the
infinitesimal transformations it is clear that ${\bf P} = (P_x,P_y)$
transforms like a vector under rotations, i.e. $\tilde {\bf P} = {\bf
  R} \cdot {\bf P}$, where ${\bf R}$ is the rotation matrix defined in
eq.~(\ref{Rmat}). It follows that the constant $\lambda$ in
eq.~(\ref{EOMlambda}) can indeed be identified with the conserved
momentum in the $y$-direction.

Using the above results it is easy to compute the change in the
orbital angular momentum $L$ under infinitesimal rotations:
\beqs
\delta L(x,\epsilon_\theta) 
&=& x \cdot\delta P_y(x,\epsilon_\theta) - 
 \left[\delta y \cdot P_x(x,\epsilon_\theta)
 + y \cdot \delta P_x(x,\epsilon_\theta) \right] 
\nonumber \\
&=& - y (q P_x - P_y) \epsilon_\theta
\eeqs
Note that this is the change at a {\it fixed} $x$, so of course there
is no variation with respect to~$x$. The change in the spin $S$ under
rotations is given by
\beqs
\delta S(x,\epsilon_\theta) 
&=& {\partial S \over \partial q}  \delta q(x,\epsilon_\theta) + 
    {\partial S \over \partial q'} \delta q'(x,\epsilon_\theta)
\nonumber \\
&=& \nu (\nu-1) 
    {y |q'|^{\nu-2} \over \left(1+q^2\right)^{3\nu-1 \over 2}}
    \left[3 q (q')^2 - (1+q^2) q'' \right] \epsilon_\theta 
\nonumber \\
&=& y (q P_x - P_y) \epsilon_\theta \,.
\eeqs
It follows that the changes in $L$ and $S$ exactly cancel each other
such that the total angular momentum $J$ does not change under
rotations, i.e.:
\beq
\delta J(x,\epsilon_\theta) =
\delta L(x,\epsilon_\theta) +
\delta S(x,\epsilon_\theta) = 0 \,.
\eeq

For completeness let us also compute the changes of the three
conserved charges under infinitesimal translations. Since the momenta
$P_x$ and $P_y$ do not depend explicitly on $y(x)$ they are trivially
invariant under translations in the $y$-direction by $\delta
y(x,\epsilon_x)$, due to eq.~(\ref{delq-transly}).  Infinitesimal
changes in the $x$-directions induce $\delta y(x,\epsilon_x)$ as given
in eq.~(\ref{dely-translx}) implying that the change of the $n^{\rm
  th}$ derivative of $y(x)$ is given by
$y^{(n)}(x,\epsilon_x)=-\epsilon_x y^{(n+1)}(x)$,
c.f.~eq.~(\ref{delq-translx}).  Then the changes in the momenta are
proportional to their total derivative, which vanishes, i.e.
\beq \label{delPtransl}
\delta P_i(x,\epsilon_x) = 
-\left({\partial P_i \over \partial q}   q' +
       {\partial P_i \over \partial q'}  q'' +
       {\partial P_i \over \partial q''} q''' \right)
\epsilon_x = 
- {dP_i \over dx} \epsilon_x = 0 \,.
\eeq
Finally the total angular momentum $J$ in eq.~(\ref{J}) does depend
explicitly on $y(x)$ and consequently changes under translations in
the $y$-direction by
\beq \label{delJ-transly}
\delta J(x,\epsilon_y) = - P_x \epsilon_y \,.
\eeq
For the change in $J$ due to an infinitesimal translation in the
$x$-direction we obtain an expression similar to the one in
eq.~(\ref{delPtransl}):
\beq \label{delJ-translx}
\delta J(x,\epsilon_x) = 
-\left({\partial J \over \partial y}   q +
       {\partial J \over \partial q}   q' +
       {\partial J \over \partial q'}  q'' +
       {\partial J \over \partial q''} q''' \right)
\epsilon_x = 
- \left({dJ \over dx} - {\partial J \over \partial x} \right)
\epsilon_x =
P_y \epsilon_x \,,
\eeq
where we used the fact that $dJ/dx=0$ and that $J$ depends explicitly
on $x$. We note that the results in eqs.~(\ref{delJ-translx})
and~(\ref{delJ-transly}) are also correct for finite translations,
since the changes do not depend on $J$.

%=========================================
\section{Scaling and dimensional analysis}
\label{scale}
%=========================================

Consider a {\it scaling transformation} that changes $x$ and $y$ by a
fraction $a$ of their original value, i.e.
\beq
\label{scaling}
\pmatrix{x \cr y} \to  
\pmatrix{\tilde x \cr \tilde y} = 
 \pmatrix{x \cr y} + a \pmatrix{x \cr y} \,.
\eeq
The infinitesimal change in $y(x)$ at a fixed $x$ under such a
transformation for $a = \epsilon_a \ll 1$ is given by
\beq
\delta y(x,\epsilon_a) = 
\left(y - \left.{\partial y \over \partial x}\right|_x  x \right) 
 \, \epsilon_a \,.
\eeq
Consequently the derivatives of $y(x)$ change by
\beqs \label{delq-scale}
\delta q(x,\epsilon_a)   &=& 
 - (q' x) \, \epsilon_a \,, \nonumber \\
\delta q'(x,\epsilon_a)  &=& 
 - (q'' x + q') \, \epsilon_a \,, \\
\delta q^{(n)}(x,\epsilon_a) &=& 
 - (q^{(n+1)} x + n q^{(n)}) \, \epsilon_a \nonumber \,.
\eeqs
Therefore it follows that some arbitrary function $A$ of $y(x)$ and
its derivatives changes under an infinitesimal scaling transformation
by
\beq \label{scalingA}
\delta A(x,\epsilon_a) =
-\left[\left({dA \over dx} - {\partial A \over \partial x} \right) x
-{\partial A \over \partial y} y 
+\sum_n {\partial A \over \partial q^{(n)}} \, n q^{(n)} 
\right] \, \epsilon_a \,.
\eeq
Using this formula it is easy to show that the conserved charges
$P_x$, $P_y$ and $J$ transform as follows under the scaling
transformation:
\beqs
\delta P_i(x,\epsilon_a) &=& -\nu P_i \, \epsilon_a~~(i=x,y) \,, \\
\delta J(x,\epsilon_a)   &=& (1-\nu) J \, \epsilon_a \,.
\eeqs
Note that for each charge the infinitesimal change is proportional to
its original value. This implies that the finite scaling
transformations are given by $P_i \to P_i \cdot \exp(-\nu a)$ and $J
\to J \cdot \exp[(1-\nu) a]$.

The interpretation of the proportionality factor follows from the
following argument.  Any quantity $A$ can be written as a product of a
dimensionless number and the dimension $[A]$. Since the problem we
discuss is purely geometrical we only have a fundamental length scale
$l_0$. So $[A]=l_0^\mu$ can be written as some power of this scale.
Now the scaling transformation in eq.~(\ref{scaling}) can be viewed as
a change of the length scale $l_0$ by some fraction $\delta l_0 =
\epsilon_a l_0$. Then to first order the corresponding change in $A$
is given by
\beq
\delta A = {\partial A \over \partial l_0} \delta l_0 
 = \mu A \, \epsilon_a \,.
\eeq
Thus the proportionality factor is nothing more than the dimension of
the quantity and we find that $[P_x]=[P_y]=l_0^{-\nu}$ and
$[J]=l_0^{1-\nu}$. It is reassuring that these results follow also
from dimensional analysis of the definitions of the conserved charges
by noting that $[x]=[y]=l_0$ and $[q^{(n)}]=l_0^{-n}$.  The dimension
of the action $[{\cal S}]=l_0^{1-\nu}$ allows us to understand {\it a
  posteriori} why the regime $0< \nu \le 1$ had to be excluded from
our analysis. If $\nu=1$ the action is invariant under scaling
transformations. In particular, we can shrink or magnify sections of
any possible solution and thereby transform it to any arbitrary shape
without changing its action. This explains why for $\nu=1$ the action
only depends on $\Delta\alpha=\alpha_f-\alpha_i$, as mentioned after
eq.~(\ref{action}). The physics jargon is to say that the action
becomes ``soft'' when $\nu$ approaches unity and it is ``critical'' at
$\nu=1$. Now if $0<\nu<1$ one can always find a ``trivial solution''
which is defined as follows: just follow the rays at $P_i$ and $P_f$
to the point where they intersect and bend the curve in the
infinitesimal vicinity of the intersection point by $\Delta\alpha$.
For this curve the action vanishes, since any length element $ds(x)$
of the straight part of the curve, where $r(x) = \infty$, does not
contribute to the action as long as $\nu>0$. Moreover the
(infinitesimal) part of the curve that is bended also does not affect
the action, because from the scaling property of the action we know
that it decreases when shrinking the unit length $l_0$ as long as
$\nu<1$. Consequently continuously scaling down the region where the
bending takes place, we achieve a zero, and hence minimal action. For
$\nu<0$ the fact that any finite straight piece of the curve gives an
infinite contribution to the action precludes this solution and for
$\nu>1$ it is not viable since, due to the inverse scaling, the action
blows up if the curve is bended strongly on a section of small length.

The fact that the momenta and the action have somewhat unusual (and
$\nu$ dependent) dimensions could easily be remedied by multiplying
the Lagrangian in eq.~(\ref{Lagrangian}) by $l_0^{\nu-1}$. Then the
action and the angular momenta would be dimensionless and the linear
momenta would have dimensions of $l_0^{-1}$.

Using eq.~(\ref{scalingA}) or just applying dimensional arguments it
follows that a scaling transformation on the right-hand side of the
EOM in eq.~(\ref{EOM}) results in a multiplication by a factor
$[1+(1-\nu) \epsilon_a]$. However since the left-hand side is zero it
is clear that the EOM is unchanged in the new coordinate system.  The
important point to note is that even though the EOM is invariant under
scaling, the transformation in eq.~(\ref{scaling}) is {\it not} a
symmetry transformation. The reason is that the change in the
Lagrangian under an infinitesimal scaling transformation,
\beq
\delta {\cal L}(x,\epsilon_a) =
-\left[{d \over dx} (x {\cal L}) 
+{\cal L}(1-\nu) \right] \, \epsilon_a \,,
\eeq
cannot be written as a total derivative for $\nu \ne 1$.  As a
consequence there is no conserved charge related to scaling.

%=====================
\section{The solution}
%=====================

The coordinate transformations discussed in the previous section are
very useful for actually solving our problem: If we manage to find the
SCF in some convenient coordinate system $\Sigma$, we can use
translations, rotations and scaling to fit the particular solution to
any boundary conditions given in some other coordinate system $\tilde
\Sigma$. Let us consider a particular solution for which $P_x=J=0$.
Then from eqs.~(\ref{Px}) and~(\ref{J}) it follows that
\beqs
0 &=& {{(\nu-1) |q'|^\nu} \over 
       {\left(1+q^2 \right)^{3\nu-1 \over 2}}} - q \, P_y \,, 
\label{zeroPx} \\
0 &=& x P_y-{\nu \sigma |q'|^{\nu-1} \over 
             \left(1+q^2\right)^{3\nu-3 \over 2}} \,.
\label{zeroJ}
\eeqs
From eq.~(\ref{zeroJ}) we get
\beq \label{qprime}
|q'| = \left({\sigma x P_y \over \nu}\right)^{1 \over \nu-1} 
 (1+q^2)^{3/2} \,,
\eeq
provided that $\sigma x P_y/\nu$ is non-negative. This requirement
implies that when $x$ changes sign also $q'$ has to change its sign
$\sigma$. Plugging the result for $|q'|$ into eq.~(\ref{zeroPx}) and
solving for $q^2$ we find
\beq \label{qsolution}
q^2 = {C \, |x|^{2\nu \over \nu-1} \over 
       1 - C \, |x|^{2\nu \over \nu-1}} \,,
\eeq
where 
%
%\beq
$
C \equiv (\nu-1)^2 |\nu|^{2\nu \over 1-\nu} |P_y|^{2 \over \nu-1}
$
%\eeq
%
can be set to unity by choosing $P_y=\pm |\nu|^\nu/|\nu-1|^{\nu-1}$.  
Because $q$ can be positive or negative there are two solutions
\beq \label{solution}
y(x)_{\pm} = \pm \int_{x_0}^x \sqrt{(\bar x^2)^{\nu \over \nu-1} \over 
                                  1-(\bar x^2)^{\nu \over \nu-1}} d\bar x 
+y_\pm(x_0) \,,
\eeq
where $y_\pm(x_0)$ are the respective integration constants.  Since
either $\nu \le 0$ or $\nu>1$ the solution in eq.~(\ref{solution}) is
defined only in the interval $-1 \le |x| < 1$. 

In the limit where $|\nu| \to \infty$ we can solve
eq.~(\ref{solution}) analytically:
\beq \label{nu-inf}
\lim_{|\nu| \to \infty} y(x)_{\pm} = 
\pm \int_{x_0}^x \sqrt{\bar x^2 \over 1-\bar x^2} d\bar x =
\pm \left(\sqrt{1-x_0^2}-\sqrt{1-x^2}\right) + y_\pm(x_0) \,.
\eeq
We see that in this case the SCF describes a segment of a circle,
which has a constant curvature radius and therefore presents the best
solution if we only care about minimal curvature along the curve. For
finite values of $\nu$ also the length of the curve plays a role.  The
integral can be expressed in terms of hypergeometric functions
$F(a,b;c;x)$, i.e.
\beq 
y(x)_{\pm} = \mp x \,{\rm Im}\left[F({\textstyle 
             {1-\nu \over 2\nu}, {1 \over 2};
             {1+\nu \over 2\nu}; |x|^{2\nu \over 1-\nu}})\right] +
             y_\pm(x_0) \,.
\eeq
We show the functions $y(x)_+$ (solid) and $y(x)_-$ (dashed) for
various $\nu$ in Fig.~\ref{SCFs1}.  The two branches are monotonic and
we have chosen the constants of integration $y_\pm(x_0)$ such that
they can be obtained from each other by a reflection with respect to
the $y$-axis.  (If one chooses the constants of integration such that
all curves go through $(1,0)$ the two branches are related by a
reflection with respect to the $x$-axis.)  Each SCF changes the sign
of its curvature at $x=0$. For large $|\nu|$ the SCF is very close to
the arc of a circle. The curves for $\nu>1$ are below the curve for
$|\nu| \to \infty$ and they become ``flatter'' and thus shorter for
smaller values of $\nu$. The curves for positive $\nu$ approach $x=0$
with a vanishing slope and their curvature radius diverges at $x=0$.
The closer $\nu$ is to unity the sooner the curve approaches the value
$y(0)$ when $x \to 0$ (which can be understood from the scaling
behavior of the action c.f.  section~\ref{scale}). The curves for 
$\nu<0$ all reside above the curve corresponding to $|\nu| \to
\infty$. For these curves both the slope $q(x)$ and the curvature 
radius $r(x)$ vanish at $x=0$.

The standard smooth connection functions shown in Fig.~\ref{SCFs1} are
the fundamental solutions to the boundary problem we want to solve.
Any specific solution consists of a segment of a standard smooth
connection function that can be viewed as a template which may be
rotated, translated and scaled in order to fit the boundary
conditions. Before we continue to describe in detail how this can be
done, it is useful to extend the SCF beyond the interval they are
defined on.

To this end we note that from eq.~(\ref{qprime}) it follows that the
curvature radius, defined in eq.~(\ref{curvature}), is given by
\beq
r(x)_\pm = |x|^{1 \over 1-\nu} \cdot {\nu-1 \over \nu} \,.
\eeq
It has the same value for the two branches of the SCF in
eq.~(\ref{solution}). Therefore, even though $q(x)$ diverges at
$|x|=1$ the curvature radius has a well-defined limit for $|x| \to 1$,
\beq
\lim_{|x| \to 1} r(x)_{\pm} = {\nu-1 \over \nu} \,.
\eeq
Thus it is natural to connect the two solutions $y(x)_{\pm}$ to one
single curve and for the following we shall fix the constant of
integration such that $\lim_{|x| \to 1} y(x)=0$ for all curves.  The
problem is that these curves are not single-valued.  In order to
express them as a single function we simply exchange the coordinates
$x$ and $y$. The resulting curves are shown in Fig.~\ref{SCFs2}. All
curves have been rescaled such that they are defined in the interval
$-1<x<1$. The interval $-1/2<x<1/2$ corresponds to the SCFs of
Fig.~\ref{SCFs1}. The continuations beyond $x=\pm 1/2$ use a segment
of the other branch, which is attached to the centerpiece such that
the curvature radius is continuous at $x=\pm 1/2$.

Now that we have these ``extended standard smooth connection
functions'' of a given $\nu>1$ for a particular set of values for the
conserved charges, namely
\beq \label{charges}
P_x = J = 0 \andthis P_y=\pm {|\nu|^\nu \over |\nu-1|^{\nu-1}} \,,
\eeq
it is not difficult to obtain a specific SCF for {\it any}\/ given
boundary conditions in eqs.~(\ref{boundary1}) and~(\ref{boundary2}).
The basic idea is to find first two points $\tilde P_i$ and $\tilde
P_f$ on a ``standard SCF'',  where the slopes correspond to the required
angles $\alpha_i$ and $\alpha_f$, and then to apply a set of
coordinate transformations to the curve in order to match the boundary
conditions.  The first step implies that we have to check whether it
is possible to find positions $\tilde x_i$ and $\tilde x_f$ somewhere
on the standard SCF such that
\beq \label{boundary3}
q(\tilde x_i) = \tan(\alpha_i+\tilde \alpha_0) ~\andthis \,
q(\tilde x_f) = \tan(\alpha_f+\tilde \alpha_0) \,, 
\eeq
where
\beq \label{alpha0}
\tan \tilde \alpha_0 = t_0(\tilde x_i,\tilde x_f) \equiv 
 {\tilde y(\tilde x_f)-\tilde y(\tilde x_i) \over 
  \tilde x_f-\tilde x_i} \,.
\eeq
Using that
$\tan(\alpha+\beta)=(\tan\alpha+\tan\beta)/(1-\tan\alpha\tan\beta)$ we
can rewrite these conditions as two coupled equations
\beqs
\tan(\alpha_i) &=& t_i(\tilde x_i,\tilde x_f) \equiv 
 {q(\tilde x_i) - t_0(\tilde x_i,\tilde x_f) \over 1 + 
  q(\tilde x_i) \, t_0(\tilde x_i,\tilde x_f)} \,, \\
\tan(\alpha_f) &=& t_f(\tilde x_i,\tilde x_f) \equiv
 {q(\tilde x_f) - t_0(\tilde x_i,\tilde x_f) \over 1 + 
  q(\tilde x_f) \, t_0(\tilde x_i,\tilde x_f)} \,.
\eeqs
This set of equations can be solved numerically.  For example one can
apply Newton's method and use the iteration scheme
\beq
\pmatrix{\tilde x_i^{(n+1)}   \cr \tilde x_f^{(n+1)}} =
\pmatrix{\tilde x_i^{(n)} \cr \tilde x_f^{(n)}} -
\left.
\pmatrix{{\partial t_i \over \partial x_i} &
         {\partial t_f \over \partial x_i} \cr
         {\partial t_i \over \partial x_f} &
         {\partial t_f \over \partial x_i} \cr}^{-1} 
\right|_{\tilde x_{i,f}^{(n)}} 
\pmatrix{t_i(\tilde x_{i,f}^{(n)})-\tan(\alpha_i) \cr 
         t_f(\tilde x_{i,f}^{(n)})-\tan(\alpha_f) } \,.
\eeq
Of course such an iterative procedure will only converge provided that
for a given $\nu$ one can indeed find two points on the standard SCF
that satisfy eq.~(\ref{boundary3}). The important observation is that
using the {\it extended\/} standard SCFs (as shown in Fig.~\ref
{SCFs2}) it is possible to find a suitable segment of the curves for
{\it any\/} given pair of $(\alpha_i,\alpha_f)$. Thus we will use
these curves in the following.

\begin{figure}[\figpos]
\begin{center}
  \mbox{\epsfig{figure=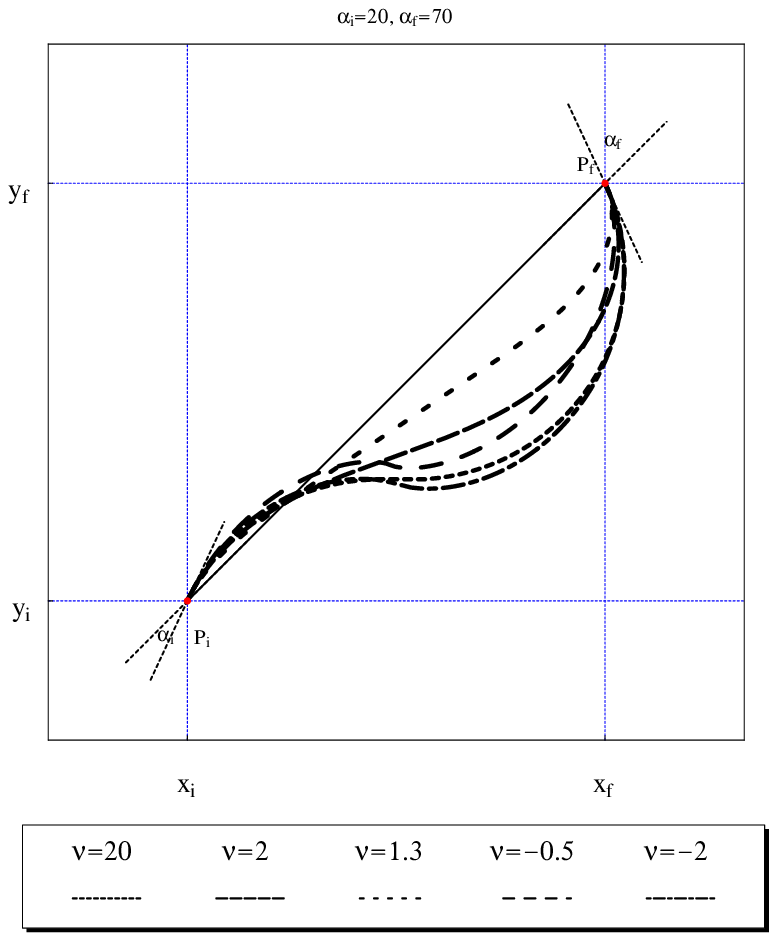,angle=0,width=17cm,height=17cm}}
\end{center}
\vspace{1cm}
\caption{Example for a specific set of smooth connection 
  functions that all satisfy the same boundary conditions. Each curve
  corresponds to a different $\nu$ as indicated in the legend. All the
  solutions consists of a segment of an extended standard smooth
  connection function that can be viewed as a template which may be
  rotated, translated and scaled in order to fit the boundary
  conditions. This example corresponds to $\alpha_i=20^\circ$ and
  $\alpha_f=70^\circ$.}
\label{SCFs} 
\end{figure}

\begin{figure}[\figpos]
\begin{center}
  \mbox{\epsfig{figure=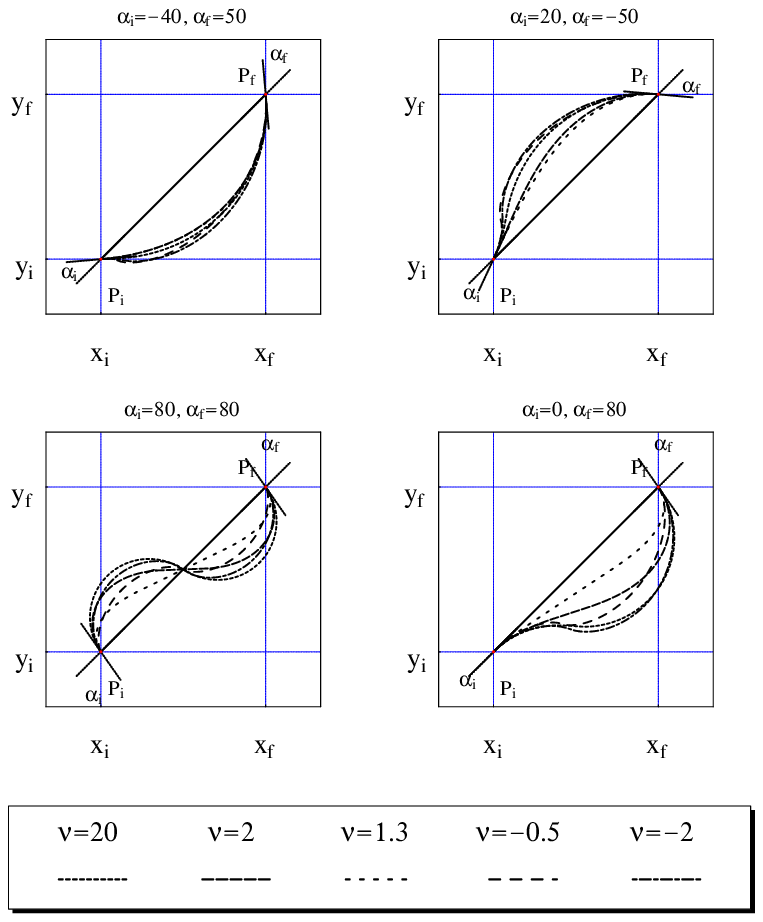,angle=0,width=17cm,height=17cm}}
\end{center}
\vspace{1cm}
\caption{Examples for solving specific choices of
  $(\alpha_i,\alpha_f)$ with the (extended) standard SCFs. Each plot
  corresponds to a particular choice of $(\alpha_i,\alpha_f)$ and
  shows the behavior of the solution for five different values of
  $\nu$ as indicated in the legend.}
\label{MoreSCFs} 
\end{figure}

In general the endpoints $\tilde P_i$ and $\tilde P_f$ of the fitting
segments do not satisfy the boundary conditions in
eqs.~(\ref{boundary1}) and~(\ref{boundary2}).  However, it is
important to realize that neither translations, nor rotations, nor
scaling transformations change the angles $\alpha_i$ and $\alpha_f$.
This is because they are defined relative to the line through $\tilde
P_i$ and $\tilde P_f$.  The only variable in eq.~(\ref{boundary3})
that does change under coordinate transformations is the angle $\tilde
\alpha_0$ that defines the direction of this line with respect to the
$x$-axis.  Thus we can apply a rotation in order to match $\tilde
\alpha_0$ with any given value in eq.~(\ref{boundary2}) for
$\alpha_0$. Since $\tilde \alpha_0$ is defined as the ratio between
the differences of the coordinates of $\tilde P_i$ and $\tilde P_f$
[see eq.~(\ref{alpha0})] it is invariant under translations and
scaling transformations. This enables us to satisfy also the boundary
conditions in eq.~(\ref{boundary1}) by first scaling the (rotated) SCF
$\tilde y(\tilde x)$ such that the distance between $\tilde P_i$ and
$\tilde P_f$ coincides with the distance between $P_i$ and $P_f$ and
then translating the resulting curve such that it connects these
points. The SCF obtained like this satisfies both
eq.~(\ref{boundary1}) and eq.~(\ref{boundary2}).  In Fig.~\ref{SCFs}
we show a specific set smooth connection functions, each corresponding
to a different $\nu$, that all satisfy the same boundary conditions.
Various other examples of smooth connection functions, that all have
been obtained by the numerical recipe described above, are shown in
Fig.~\ref{MoreSCFs}. Each plot corresponds to a particular choice of
$(\alpha_i,\alpha_f)$ and shows the behavior of the solution for five
different values of $\nu$.

Finally we would like to compare the SCF discussed in this paper with
other ``conventional'' interpolation functions. In particular for
practical purposes it is important to know how much better the optimal
path (i.e. the SCF) is with respect to some approximation.  To this
end it is useful to have a closed expression for the action in
eq.~(\ref{action}), which determines the ``smoothness'' of any curve
$y(x)$. Using eq.~(\ref{Px}) its integrand can be written in terms of
the linear momenta as
\beq
\tilde{\cal L} = {P_x + P_y \, y'(x) \over \nu-1} \,.
\eeq
If $y(x)$ solves the EOM, then $P_x$ and $P_y$ are constants and
$\tilde{\cal L}$ can easily be integrated resulting in
\beq
\tilde{\cal S} = {P_x (x_f-x_i) + P_y (y_f-y_i) \over \nu-1} \,.
\eeq
Note that $\tilde{\cal S}$ is manifestly invariant under rotations and
translations. If $y(x)$ does not solve the EOM, the integration in
eq.~(\ref{action}) in general has to be done numerically. Standard
interpolation curves are often given in parametric form, i.e. as
$\{x(t),y(t)\}$, where $t$ varies within a given interval $[t_i,t_f]$
along the curve. In this case the action is given by
\beq \label{actionpara}
\tilde{\cal S}_p = 
\int_{t_i}^{t_f} \! 
 {|\dot{x} \ddot{y} - \ddot{x} \dot{y}|^\nu \over 
  (\dot{x}^2 + \dot{y}^2)^{3\nu-1 \over 2}} \, dt \,, 
\eeq
where the dot denotes a derivative with respect to $t$.  We note that
in principle one could also minimize this action in order to determine
the smooth connection functions. However in this case the
Euler-Lagrange formalism results in two coupled non-trivial
differential equations, which appear to be even harder to solve than
the EOM we got in eq.~(\ref{EOM}). Using eq.~(\ref{actionpara}) we
have computed the weight functional for cubic Bezier curves. In
general for small $\nu$ the SCF can be well-approximated by an
appropriate Bezier curve and the respective values for the action only
differ at most by a few percent.  However, for larger $\nu$ it is
increasingly difficult to match the SCF by a Bezier curve and the
difference between the actions becomes significant. This is because
for $\nu \gg 1$ the SCF approaches an arc of a circle which cannot be
parameterized by polynomial functions.  Thus for applications where
speed is more important than accuracy the Bezier curves are the
favorite solution, but whenever accuracy is crucial or an exact
solution is called for then the smooth connection functions are
needed.

%================================
\section{Conclusions and Outlook}
%================================

We have presented a generic solution of how to connect two points in a
plane by a smooth curve that goes through these points with a given
slope. Our approach uses extensively notions that are well-known in
classical mechanics. The smooth connection function $y(x)$ has to
satisfy certain boundary conditions and to minimize the action
functional that reflects the smoothness of any function by integrating
over the inverse curvature radius (to some power $\nu$) times the
length element along the curve.  Minimizing the action via the
Euler-Lagrange formalism leads to the equation of motion which is a
complicated non-linear third order differential equation. However the
translational and rotational symmetries of the problem are of great
help, since they imply conserved charges, i.e. the linear and the
angular momenta, that help to simplify the problem significantly.
Making a specific choice for the charges it is possible to obtain the
SCF $y(x)$ for a given $\nu$ in terms of hypergeometric functions.
Applying the appropriate coordinate transformation to this solution
allows to adjust the SCF to arbitrary boundary conditions.

We have worked out in detail the basic formalism to find explicitly
the smoothest connection between two points in the two dimensional
Euclidean space. Several generalizations and extensions of this basic
problems are possible: First, the number of points that define the SCF
can be increased. For a single curve the solution will still be
determined by four boundary conditions, but one may choose different
conditions than those in eqs.~(\ref{boundary1}) and~(\ref{boundary2}).
Also it is possible to paste together several elementary solutions in
order to find interpolations between several points which cannot be
achieved by a single SCF. How to do this best gives rise to a new
optimization problem. Finally we note that one can also choose to
apply our formalism to a different geometry.  For example one may
consider a time-dependent SCF $x(t)$ in Minkowski space, where the
metric is defined via $r^2=x^2-t^2$. Also extensions of the problem to
higher dimensional spaces are conceivable. In this case the SCF would
describe some smooth manifold that connect two extended objects (like
strings).

%%%%%%%%%%%
% Figures %
%%%%%%%%%%%

\putFig{SCF}{Different ``Smooth Connection Functions'' that connect
  the initial point $P_i: (x_i,y_i)$ with the final point $P_f:
  (x_f,y_f)$. At $P_i$ and $P_f$ the curves are tangent to the
  associated rays (dashed) which are specified by their inclination
  angles, $\alpha_i$ and $\alpha_f$, with respect to the vector
  pointing from $P_i$ to $P_f$. The light-gray curve (resulting from
  $\nu=1.2$) is the shortest, but has the strongest curvature.
  Conversely, the dark-grey curve ($\nu=20$) is the longest, but
  almost corresponds to an arc of a circle with minimal curvature. The
  intermediate grey curve ($\nu=2$) presents a more balanced
  compromise between minimal length and curvature.}{13}{13}{8.1}

\putFig{SCFs1}{Standard smooth connection functions for various $\nu$.
  The two branches are monotonic and we have chosen the constants of
  integration such that they can be obtained from each other by a
  reflection with respect to the $y$-axis. Each SCF changes the sign
  of its curvature at $x=0$. For large $|\nu|$ the SCF is very close
  to the arc of a circle. The curves for $\nu>1$ are below the curve
  corresponding to $|\nu| \to \infty$. These curves become ``flatter''
  and thus shorter for smaller values of $\nu$. They approach $x=0$
  with a vanishing slope and their curvature radius diverges at $x=0$.
  The closer $\nu$ is to unity the sooner the curve approaches the
  value $y(0)$ when $x \to 0$. The curves for $\nu<0$ all reside above
  the curve corresponding to $|\nu| \to \infty$. For these curves both
  the slope $q(x)$ and the curvature radius $r(x)$ vanish at
  $x=0$.}{13}{13}{7.7}

\putFig{SCFs2}{Extended standard smooth connection functions for
  various $\nu$.  These functions are obtained from the SCFs of
  Fig.~\ref{SCFs1} by exchanging the coordinates $x$ and $y$. All
  curves have been rescaled such that they are defined in the interval
  $-1<x<1$. The interval $-1/2<x<1/2$ corresponds to the SCFs of
  Fig.~\ref{SCFs1}. The continuations beyond $x=\pm 1/2$ use a segment
  of the other branch, which is attached to the centerpiece such that
  the curvature radius is continuous at $x=\pm 1/2$.}{14}{14}{7.7}

\newpage

\newpage

\end{document}